# Eigenspace-Based Minimum Variance Combined with Delay Multiply and Sum Beamformer: Application to Linear-Array Photoacoustic Imaging

Moein Mozaffarzadeh, *Member, IEEE,* Ali Mahloojifar*, *Member, IEEE,* Vijitha Periyasamy, *Member, IEEE,* Manojit Pramanik, and Mahdi Orooji, *Member, IEEE*

*Abstract*—In Photoacoustic imaging, Delay-and-Sum (DAS) algorithm is the most commonly used beamformer. However, it leads to a low resolution and high level of sidelobes. Delay-Multiply-and-Sum (DMAS) was introduced to provide lower sidelobes compared to DAS. In this paper, to improve the resolution and sidelobes of DMAS, a novel beamformer is introduced using Eigenspace-Based Minimum Variance (EIBMV) method combined with DMAS, namely EIBMV-DMAS. It is shown that expanding the DMAS algebra leads to several terms which can be interpreted as DAS. Using the EIBMV adaptive beamforming instead of the existing DAS (inside the DMAS algebra expansion) is proposed to improve the image quality. EIBMV-DMAS is evaluated numerically and experimentally. It is shown that EIBMV-DMAS outperforms DAS, DMAS and EIBMV in terms of resolution and sidelobes. In particular, at the depth of 11 $mm$ of the experimental images, EIBMV-DMAS results in about 113 $dB$ and 50 $dB$ sidelobe reduction, compared to DMAS and EIBMV, respectively. At the depth of 7 $mm$, for the experimental images, the quantitative results indicate that EIBMV-DMAS leads to improvement in Signal-to-Noise Ratio (SNR) of about 75% and 34%, compared to DMAS and EIBMV, respectively.

*Index Terms*—Photoacoustic imaging, beamforming, Delay-Multiply-and-Sum, Eigenspace-based minimum variance, linear-array imaging.

## I. INTRODUCTION

PHOTOACOUSTIC imaging (PAI) is an emerging medical imaging modality, having merits of the optical imaging contrast and the Ultrasound (US) imaging spatial resolution [1]–[3]. In PAI, a short electromagnetic pulse is used to induce US waves, based on the thermoelastic effect [4], [5]. PAI has been used in various field of studies [6]–[8]. In 2002, for the first time, Photoacoustic Tomography (PAT) was successfully used as *in vivo* functional and structural brain imaging modality in small animals [9]. Recently, low-cost PAT systems are extensively being investigated [10], [11]. One of the challenging matters in PAI is image reconstruction. Imperfect reconstruction algorithms cause the images having inherent artifacts [12]. Having these artifacts mitigated would significantly improve the Photoacoustic (PA) image quality.

*Corresponding author
M. Mozaffarzadeh, A. Mahloojifar and M. Orooji are with the Department of Biomedical Engineering, Tarbiat Modares University, Tehran, Iran (e-mail: moein.mfh@modares.ac.ir; mahlooji@modares.ac.ir; morooji@modares.ac.ir).
V. Periyasamy and M. Pramanik are with the Nanyang Technological University, School of Chemical and Biomedical Engineering, 62 Nanyang Drive, Singapore (e-mail: vijitha@ntu.edu.sg; manojit@ntu.edu.sg).

In PAI, both contrast and resolution are important, and it has been investigated in the different publications [1], [3]. Usually, enhancing one of them would lead to degrade the other one. Having a good spatial resolution along with a high contrast has not been extensively investigated. Thus, it is worth to develop an algorithm which provides these properties together. Beamforming algorithms used in US imaging can be used in linear-array PAI as a result of high similarity between US and PA detected signals. Delay-And-Sum (DAS) and Minimum variance (MV) are two common beamforming methods [13], [14]. DAS is the most common beamforming method in linear-array PAI due to its simple implementation. However, it is a blind beamformer, leading to a wide mainlobe and high level of sidelobes [13], [15], [16]. Matrone *et al.* proposed a new algorithm namely Delay-Multiply-and-Sum (DMAS) as a beamforming technique, used in medical US imaging [17]. Recently, we introduced a novel beamforming algorithm, outperforming DMAS in terms of the contrast and sidelobes [18], [19]. In addition, for linear-array PAI, MV was combined with DMAS to improve the resolution of the DMAS while the sidelobes are retained [15], [20], [21]. Two modifications of Coherence Factor (CF) have been introduced for linear-array PAI, in order to have a lower sidelobes and higher resolution compared to the conventional CF [22], [23]. In this paper, a novel beamforming algorithm, namely Eigenspace-Based Minimum Variance-DMAS (EIBMV-DMAS), is introduced. DMAS algebra is expanded, and it is shown that in each term of the expansion, there is a DAS algebra. Since the DAS leads to low resolution images, we proposed to combine EIBMV with the expansion of DMAS algebra, which necessitates some modifications discussed in the paper. A preliminary version of this work has been already reported in [21]. However, in this paper, we present a more completed description of this approach and evaluate its performance and the effects of parameters numerically and experimentally. It is shown that using EIBMV-DMAS leads to resolution improvement and sidelobe levels reduction, at the expense of higher computational burden, in comparison with DMAS and EIBMV, respectively.

The rest of the paper is organized as follows. In the section II, the DMAS and EIBMV beamforming algorithms, along with the proposed method, are presented. Numerical and experimental results are presented in the section III and IV, respectively. The advantages and disadvantages of the proposed method are discussed in the section V, and finally, conclusion is presented in the section VI.

## II. METHODS AND MATERIALS

### A. Beamforming

Beamforming algorithms such as DAS can be utilized to reconstruct the image from the PA signals, detected by a linear-array of US transducer. DAS formula is as follows:

$$y_{DAS}(k) = \sum_{i=1}^{M} x_i(k - \Delta_i), \quad (1)$$

where $y_{DAS}(k)$ is the output of the beamformer, $M$ is the number of elements of array, $k$ is the time index, and $x_i(k)$ and $\Delta_i$ are the detected signals and the corresponding time delay for detector $i$, respectively. DMAS calculates corresponding sample for each element of the array, the same as DAS, but before summation, the samples are combinatorially coupled and multiplied. The DMAS formula is given by:

$$y_{DMAS}(k) = \sum_{i=1}^{M-1} \sum_{j=i+1}^{M} x_i(k - \Delta_i) x_j(k - \Delta_j). \quad (2)$$

To overcome the dimensionally squared problem of (2), following equations are suggested [17]:

$$\hat{x}_{ij}(k) = \text{sign}[x_i(k - \Delta_i) x_j(k - \Delta_j)] \sqrt{|x_i(k - \Delta_i) x_j(k - \Delta_j)|}. \quad (3)$$

$$y_{DMAS}(k) = \sum_{i=1}^{M-1} \sum_{j=i+1}^{M} \hat{x}_{ij}(k). \quad (4)$$

The procedure of DMAS algorithm can be considered as a correlation process which uses the auto-correlation of the aperture. In other words, the output of this beamformer is the spatial coherence of the PA signals, and it is a non-linear beamforming algorithm.

### B. Eigenspace-Based Minimum Variance

The output of the MV adaptive beamformer is given by:

$$y(k) = W^H(k) X_d(k) = \sum_{i=1}^{M} w_i(k) x_i(k - \Delta_i), \quad (5)$$

where $X_d(k)$ is time-delayed array detected signals $X_d(k) = [x_1(k), x_2(k), ..., x_M(k)]^T$, $W(k) = [w_1(k), w_2(k), ..., w_M(k)]^T$ is the beamformer weights, and $(.)^T$ and $(.)^H$ represent the transpose and conjugate transpose, respectively. The detected array signals can be written as follows:

$$X(k) = s(k) + i(k) + n(k) = s(k)a + i(k) + n(k), \quad (6)$$

where $i(k)$, $s(k)$ and $n(k)$ are the interference, the desired signal and noise components received by array transducer, respectively. Parameters $s(k)$ and $a$ are the signal waveform and the related steering vector, respectively. MV bemaformer adaptively weighs the calculated samples using the following equation:

$$\min_{W} W^H R_{i+n} W, \quad s.t. \quad W^H a = 1, \quad (7)$$

where $R_{i+n}$ is the $M \times M$ interference-plus-noise covariance matrix. The solution of (7) is given by [24]:

$$W_{opt} = \frac{R_{i+n}^{-1} a}{a^H R_{i+n}^{-1} a}. \quad (8)$$

In practical application, interference-plus-noise covariance matrix is unavailable. Consequently, the sample covariance matrix is used instead of unavailable covariance matrix using the N recently received samples and is given by:

$$\hat{R} = \frac{1}{N} \sum_{n=1}^{N} X_d(n) X_d(n)^H. \quad (9)$$

The diagonal loading, the spatial averaging and the temporal averaging can also be used to improve the quality of the covariance matrix estimation [14], [25]. The weights for EIBMV algorithm are generated by projecting the optimal weights for MV algorithm to a signal subspace constructed from the eigenspace of the estimated covariance matrix. Eigen decomposition of $\hat{R}_l$ can be written as follows:

$$\hat{R}_l = U \Lambda^{-1} U^H, \quad (10)$$

where $\Lambda = \text{diag}[\lambda_1, \lambda_2, ..., \lambda_L]$ in which $\lambda_1 \geq \lambda_2 \geq ... \geq \lambda_L$ are eigenvalues in the descending order, and $U = [u_1, u_2, ..., u_L]$ in which $u_i$, $i = 1, 2, ..., L$, are the orthonormal eigenvectors corresponding to $\lambda_i$, $i = 1, 2, ..., L$. $L$ is the length of sub-array used in spatial averaging. the signal subspace $E_s$ is obtained using the eigenvectors corresponding to the first largest eigenvalues, as follows:

$$E_s = [u_1, ..., u_{Num}], \quad (11)$$

where $Num$ is the number of eigenvectors. Finally, the weights for EIBMV can be generated as follows:

$$W_{EIBMV} = E_s E_s^H W_{opt}. \quad (12)$$

It should be mentioned that the eigenvectors whose related eigenvalues were larger than $\sigma$ times the largest one were used in this paper. The performance evaluation of EIBMV is presented in [26] extensively. After estimation of covariance matrix, the output of EIBMV beamformer is given by:

$$\hat{y}(k) = \frac{1}{M - L + 1} \sum_{l=1}^{M-L+1} W_{EIBMV}^H(k) X_d^l(k), \quad (13)$$

where $W_{EIBMV}(k) = [w_1(k), w_2(k), ..., w_L(k)]^T$, and $X_d^l(k) = [x_d^l(k), x_d^{l+1}(k), ..., x_d^{l+L-1}(k)]$ is the delayed input signal for the $l_{th}$ subarray [26].

### C. Proposed Method

In this paper, it is proposed to combine the EIBMV adaptive beamformer with expansion of DMAS algorithm to improve the resolution and level of sidelobes in DMAS. To illustrate this, consider the expansion of the DMAS algorithm which can be written as follows:

$$y_{DMAS}(k) = \sum_{i=1}^{M-1} \sum_{j=i+1}^{M} x_{id}(k) x_{jd}(k) =$$
$$x_{1d}(k) \underbrace{\left[ x_{2d}(k) + x_{3d}(k) + x_{4d}(k) + ... + x_{Md}(k) \right]}_{\text{first term}} \quad (14)$$
$$+ ... + \underbrace{\left[ x_{(M-1)d}(k).x_{Md}(k) \right]}_{(M-1)\text{th term}}.$$

where $x_{id}(k)$ and $x_{jd}(k)$ are delayed detected signals for element $i$ and $j$, respectively, and we hold this notation all over this section. In (14), in every terms, there exists a summation procedure which is a type of DAS algorithm. It is proposed to use EIBMV adaptive beamformer in each term instead of the DAS. To this end, we need to carry out some modifications and prepare the expansion in (14) for the proposed method. Following section contains the essential modifications.

*1) Modified DMAS:* The quality of covariance matrix estimation in EIBMV is highly affected by the selected length of subarray. $M/2$ and $1$ are considered as the upper and lower boundaries, respectively. In (14), each term can be considered as a DAS algorithm with different number of elements of array. Limited number of entries in each term causes problem for EIBMV algorithm due to the limited length of the subarray. This problem can be addressed by adding the unavailable elements in each term in order to acquire large enough number of available elements and consequently high quality covariance matrix estimation. The extra terms, needed to address the problem, are given by:

$$y_{extra}(k) = \sum_{i=M-2}^{2} \sum_{j=i-1}^{1} x_{id}(k) x_{jd}(k) + y_{extra^*}(k)$$
$$= x_{(M-2)d}(k) \left[ x_{(M-3)d}(k) + ... + x_{2d}(k) \right] + x_{1d}(k) \right] + ... \quad (15)$$
$$+ x_{3d}(k). \left[ x_{2d}(k) + x_{1d}(k) \right] + x_{2d}(k) x_{1d}(k) + y_{extra^*}(k),$$

where $y_{extra^*}(k) = x_{Md}(k) \left[ x_{(M-1)d}(k) + ... + x_{2d}(k) + x_{1d}(k) \right]$. (15) is used to make the terms in (14) ready to adopt an EIBMV algorithm. By adding (14) and (15), a modified version of DMAS algorithm namely modified DMAS (MDMAS) is obtained as follows:

$$y_{MDMAS}(k) = y_{DMAS}(k) + y_{extra}(k)$$
$$= \sum_{i=1}^{M} \sum_{j=1, j\neq i}^{M} x_{id}(k) x_{jd}(k) =$$
$$= x_{1d}(k) \underbrace{\left[ x_{2d}(k) + x_{3d}(k) + ... + x_{(M-1)d}(k) + x_{Md}(k) \right]}_{\text{first term}}$$
$$+ ... + x_{Md}(k) \underbrace{\left[ x_{1d}(k) + x_{2d}(k) + ... + x_{(M-2)d}(k) + x_{(M-1)d}(k) \right]}_{M\text{th term}}.$$
(16)

Since all the cross-products are considered twice in (16), this equation leads to images the same as DMAS. Now, the combination of MDMAS algorithm and EIBMV beamformer is mathematically satisfied. The expansion of MDMAS combined with EIBMV beamformer can be written as follows:

$$y_{MDMAS\_2}(k) = \sum_{i=1}^{M} x_{id}(k) \left( \boldsymbol{W}_{i,M-1}^{H}(k) \boldsymbol{X}_{id,M-1}(k) \right) =$$
$$\sum_{i=1}^{M} x_{id}(k) \left( \left[ \left( \sum_{j=1}^{M} w_j(k) x_{jd}(k) \right) \right] - w_i(k) x_{id}(k) \right) = \quad (17)$$
$$\sum_{i=1}^{M} x_{id}(k) \underbrace{\left( \sum_{j=1}^{M} w_j(k) x_{jd}(k) \right)}_{EIBMV} - \sum_{i=1}^{M} x_{id}(k) \left( w_i(k) x_{id}(k) \right),$$

where, in $\boldsymbol{W}_{i,M-1}$ and $\boldsymbol{X}_{id,M-1}$, the $i_{th}$ element of the array is ignored in calculation and as a result, the length of these vectors becomes $M-1$ instead of $M$. Considering (17), the expansion can be written based on a summation which is considered as a DAS algebra. To illustrate, consider following expansion:

$$y_{MDMAS\_3}(k) = \sum_{i=1}^{M} \underbrace{\left[ x_{id}(k) \underbrace{\left( \sum_{j=1}^{M} w_j(k) x_{jd}(k) \right)}_{EIBMV} - w_i(k) x_{id}^2(k) \right]}_{i_{th} term}.$$
(18)

It is proved that DAS leads to low quality images and high level of sidelobes. Obviously, in (18), expansion leads to a summation and this summation can be considered as a DAS. As the final step of EIBMV-DMAS development, it is proposed to use another EIBMV instead of DAS (having EIBMV instead of outer summation in (18)) in order to reduce the contribution of off-axis signals and noise of imaging system. EIBMV-DMAS formula can be written as follows:

$$y_{EIBMV-DMAS}(k) =$$
$$\sum_{i=1}^{M} w_{i,new} \underbrace{\left( x_{id}(k) \left( \sum_{j=1}^{M} w_j(k) x_{jd}(k) \right) - w_i(k) x_{id}^2(k) \right)}_{i_{th} term}, \quad (19)$$

where $w_{i,new}$ is the calculated weight for each term in (19) using (12) while the steering vector is a vector of ones. In the section III, it is shown that EIBMV-DMAS beamformer results in resolution improvement and sidelobes level reduction.

## III. NUMERICAL RESULTS AND PERFORMANCE ASSESSMENT

In this section, numerical results are presented to illustrate the performance of the proposed algorithm in comparison with DAS, DMAS and EIBMV.

### A. Simulated Point Target

*1) Simulation Setup:* The K-wave Matlab toolbox was used to simulate the numerical study [27]. Ten 0.1 $mm$ radius spherical absorbers as initial pressure were positioned along the vertical axis every 5 $mm$ beginning 25 $mm$ from transducer surface. Imaging region was 20 $mm$ in lateral axis and 50 $mm$ in vertical axis. A linear-array having $M$=128 elements operating at 4 $MHz$ central frequency and 77% fractional bandwidth was used to detect the PA signals generated from defined initial pressures. The speed of sound was assumed to be 1540 $m/s$ during simulations. The sampling frequency was 50 $MHz$, subarray length $L=M/2$, $K=5$ and $\Delta = 1/10L$ for all the simulations. Also, a band-pass filter was applied by a Tukey window ($\alpha$=0.5) to the beamformed signal spectra, covering 6-15 $MHz$, to pass the necessary information, generated after these non-linear operations, while keeping the one centered on $2f_0$ almost unaltered. At the end, the obtained lines are normalized and log-compressed to form the PA images. The temporal averaging for covariance matrix estimation is performed over $2K + 1$ [14].

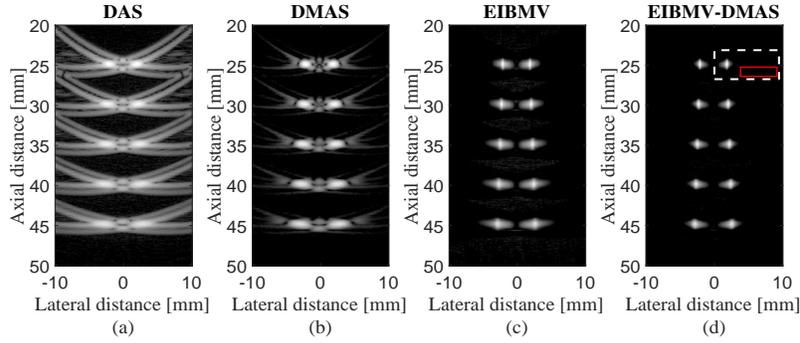

Fig. 1: Reconstructed PA images using (a) DAS, (b) DMAS, (c) EIBMV (L = M/2, K = 5, $\sigma$ = 0.7), (d) EIBMV-DMAS (L = M/2, K = 5, $\sigma$ = 0.7). All images are shown with a dynamic range of 60 $dB$. Received signals have a SNR = 50 $dB$.

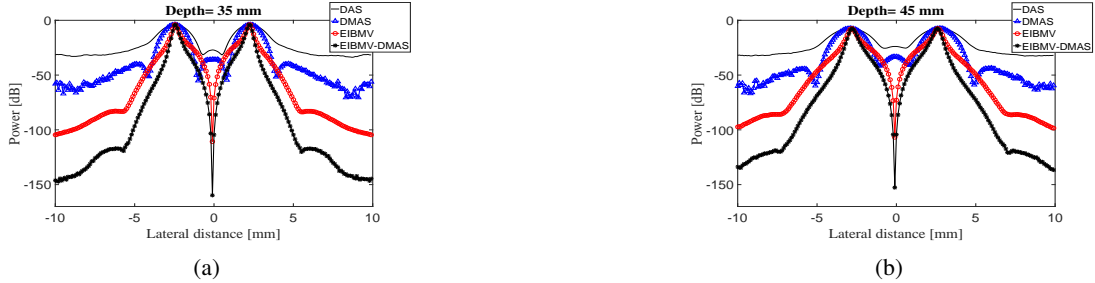

Fig. 2: Lateral variation of DAS, DMAS, EIBMV and EIBMV-DMAS at the depths of (a) 35 $mm$ and (b) 45 $mm$.

*2) Qualitative Evaluation:* Fig. 1 shows the output of DAS, DMAS, EIBMV and EIBMV-DMAS beamformers. As seen in Fig. 1, DAS leads to a low quality image having high level of sidelobes and a low resolution. DMAS reduces the level of sidelobes and improves the image quality. However, the resolution of the reconstructed image using DMAS is well improved compared to DAS. EIBMV improves the resolution, but the level of sidelobes degrade the image, as shown in Fig. 1(c). The reconstructed image using EIBMV-DMAS is shown in Fig. 1(d), and it can be seen that the level of sidelobes in EIBMV is reduced. To assess the different beamforming algorithms in details, the lateral variations of the formed images are shown in Fig. 2. The less noisy regions use the off-axis signals as an advantage to suppress noise. The lateral variations at the depth of 35 $mm$ is shown in Fig. 2(a), and it can be seen that DAS, DMAS, EIBMV and EIBMV-DMAS result in about -30 $dB$, -52 $dB$, -82 $dB$ and -120 $dB$ sidelobes level, respectively. Moreover, width of mainlobe in EIBMV-DMAS earns the narrowest shape compared to other beamformers, and the valley of the lateral variations is highly reduced. The proposed method has been evaluated at the presence of high level of noise, and the reconstructed images, along with the lateral variations, are shown in Fig. 3 and Fig. 4, respectively. As demonstrated, the presence of noise is reduced, the resolution is improved, and the sidelobes are further degraded using EIBMV-DMAS, compared to other methods.

*3) Quantitative Evaluation:* To quantitatively compare the performance of the beamformers, the full-width-half-maximum (FWHM) in -6 $dB$ and signal-to-noise ratio (SNR) are calculated in all imaging depths using point targets in the reconstructed images, presented in TABLE I. As can be seen in TABLE I, the EIBMV-DMAS results in the narrowest -6 $dB$ width of mainlobe in all imaging depths compared to other beamformers. In particular, consider depth of 40 $mm$ where FWHM for DAS, DMAS, EIBMV and EIBMV-DMAS is about 2.04 $mm$, 1.43 $mm$, 0.53 $mm$ and 0.43 $mm$, respectively. The SNRs are calculated using following equation:

$$SNR = 20\log_{10} P_{signal}/P_{noise}. \qquad (20)$$

where $P_{signal}$ and $P_{noise}$ are difference of maximum and minimum intensity of a rectangular region including a point target (white dashed rectangle in Fig. 1(d)), and standard deviation of the noisy part of the region (red rectangle in Fig. 1(d)), respectively [28]. As seen in TABLE I, EIBMV-DMAS outperforms other beamformers, having a higher SNR. Consider, in particular, depth of 40 $mm$ where SNR for DAS, DMAS, EIBMV and EIBMV-DMAS is 16.58 $dB$, 18.74 $dB$, 22.32 $dB$ and 23.72 $dB$, respectively.

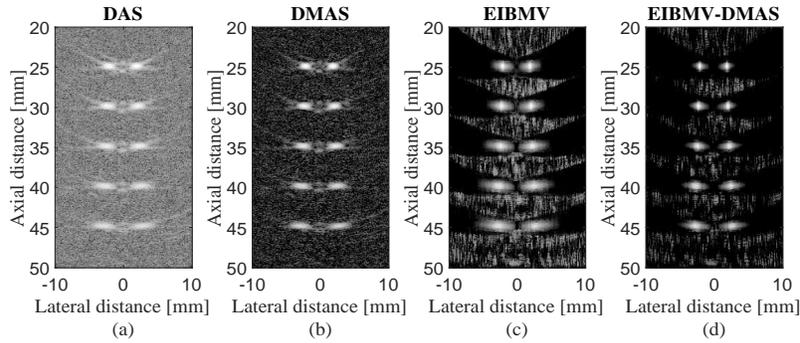

Fig. 3: Reconstructed PA images using (a) DAS, (b) DMAS, (c) EIBMV (L = M/2, K = 5, $\sigma$ = 0.7), (d) EIBMV-DMAS (L = M/2, K = 5, $\sigma$ = 0.7). All images are shown with a dynamic range of 60 $dB$. Received signals have a SNR = 15 $dB$.

TABLE I: SNR ($dB$) and FWHM ($mm$) Values at Different Depths Using Targets in Fig. 1.

| Depth($mm$) | SNR ($dB$) | | | | -6$dB$ FWHM ($mm$) | | | |
|---|---|---|---|---|---|---|---|---|
| | DAS | DMAS | EIBMV | EIBMV-DMAS | DAS | DMAS | EIBMV | EIBMV-DMAS |
| 25 | 18.89 | 20.99 | 25.85 | 26.68 | 1.22 | 0.88 | 0.24 | 0.22 |
| 30 | 18.12 | 20.34 | 24.55 | 25.33 | 1.49 | 1.06 | 0.32 | 0.29 |
| 35 | 17.26 | 19.43 | 23.34 | 24.46 | 1.78 | 1.27 | 0.45 | 0.36 |
| 40 | 16.58 | 18.74 | 22.32 | 23.72 | 2.04 | 1.43 | 0.51 | 0.43 |
| 45 | 16.05 | 18.23 | 21.81 | 23.04 | 2.36 | 1.65 | 0.59 | 0.51 |

## IV. EXPERIMENTAL RESULTS

In this section, to evaluate the EIBMV-DMAS algorithm, the results of the designed experiment are presented.

### A. Experimental Setup

The validation of the proposed algorithm was carried out on the PA data generated from four point-source targets. The PA signals were acquired using a clinical linear-array transducer. The Nd:YAG pump laser (Continuum, Surelite Ex, San Jose, California, USA) was used to generate pulses of 532 $nm$ at a pulse repetition frequency of 10 $Hz$ for excitation [29], [30]. The laser beam was diverged to illuminate the point sources. The circular beam of diameter 5 $cm$ (area of ∼ 19.63 $cm^2$) was passed through the wall of water tank to illuminate the point sources. The fluence was 10 $mJ/cm^2$ which is within the permissible limits of American Nationals Standards institute of 20 $mJ/cm^2$ [31]. The laser beam is diverged to illuminate the points source. Pencil leads (diameter 0.5 $mm$ and length 75 $mm$) were used as point target glued to a glass plate such that neither the optical fluence nor the acoustic signal from one of them mask the any other point source. The target was immersed in water for acoustic coupling. Generated PA signal was acquired using a dual-mode clinical ultrasound system (E-CUBE 12R, Alpinion, South Korea) and 128 elements linear-array transducer L3-12, which had an active area of 3.85 $cm$ × 1 $cm$ and its central frequency is 8.5 $MHz$ with 95% fractional bandwidth [32], [33]. The US system has a 64 channel parallel data acquisition. The data acquisition was triggered by the trigger from the laser. Therefore, a PA image was formed after two laser pulses. In addition, PA images were acquired at a frequency of 5 $Hz$. Axial resolution of the system was 0.2 $mm$, and the lateral resolution was 0.3 $mm$ [29]. Acquired radio frequency data were saved in the local machine and used later for testing the reconstruction algorithm.

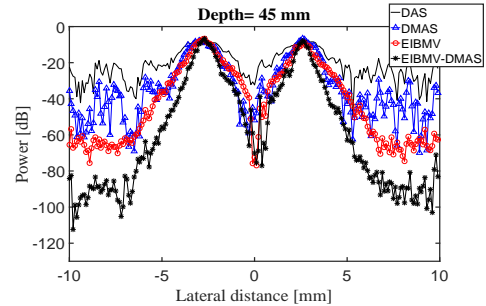

Fig. 4: Lateral variation of DAS, DMAS, EIBMV and EIBMV-DMAS at the depths of 45 $mm$.

### B. Qualitative and Quantitative Evaluation

The reconstructed images are shown in Fig. 5, along with a zoomed version in Fig. 6. We used a dynamic range of 80 $dB$ for the experimental images to show the superiority of the proposed method better. As can be seen in the reconstructed image using DAS, shown in Fig. 5(a), high level of noise and sidelobes reduce the quality of the PA image. The sidelobe levels can be clearly seen around the targets. The DMAS reduces the effects of noise and sidelobes, but the resolution is not good enough and the presence of sidelobes still degrade the image quality. EIBMV results in a high resolution image, and the targets are quite detectable. EIBMV-DMAS further reduces the sidelobes and presence of noise in the reconstructed PA image, in comparison with EIBMV. To compare the performance of the beamformers in details (using the experimental data), the lateral variations at two depths of imaging are shown in Fig. 7. As demonstrated, at both depths, the sidelobes of the proposed method are lower than other methods. Consider, for example, the depth of 11 $mm$ where DAS, DMAS, EIBMV and EIBMV-DMAS result in -35 $dB$, -47 $dB$, -110 $dB$ and -160 $dB$ level of sidelobes, respectively. As a result, EIBMV-DMAS reduces the level of sidelobes for about 113 $dB$ and 50 $dB$ compared to DMAS and EIBMV, respectively. To compare the experimental results quantitatively, SNR metrics has been utilized. The results at two different depths are presented in TABLE II. As can be seen, the SNR for EIBMV-DMAS method is higher than other beamformers.

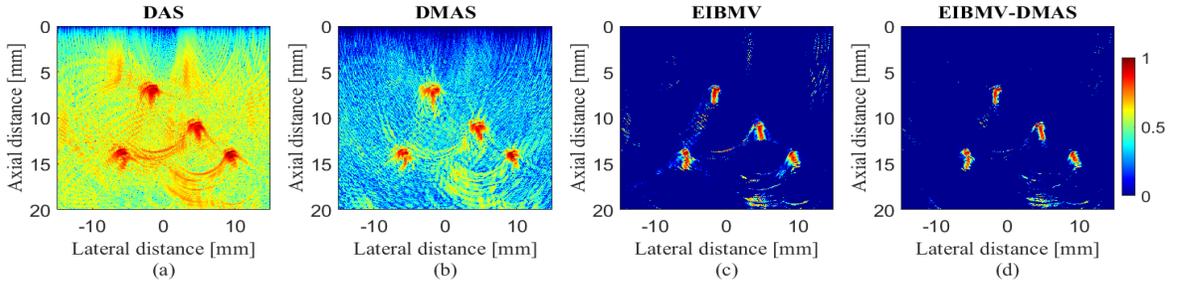

Fig. 5: Experimental reconstructed PA images. (a) DAS, (b) DMAS, (c) EIBMV (L=M/3, K=0, $\sigma$ = 0.8), (d) EIBMV-DMAS (L=M/3, K=0, $\sigma$ = 0.8). All images are shown with a dynamic range of 80 $dB$.

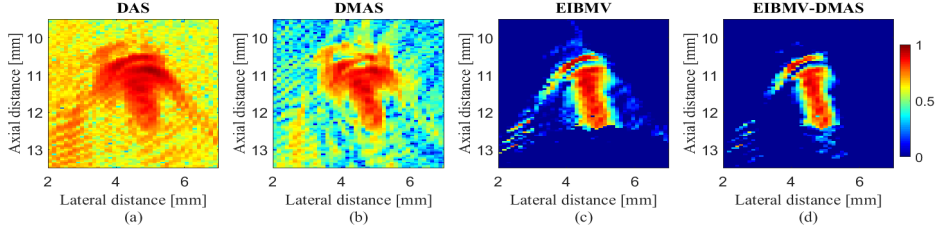

Fig. 6: The zoomed version of the PA images shown in Fig. 5 using the target located at the depth of 11 $mm$.

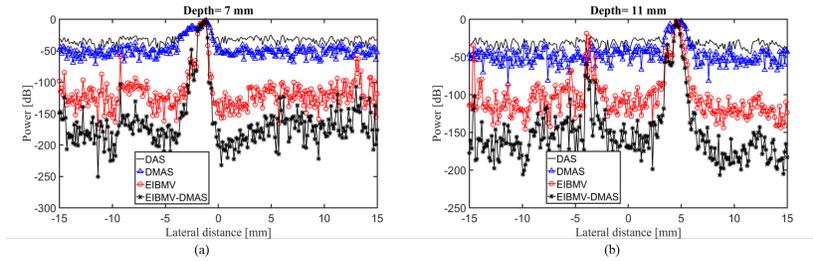

Fig. 7: Lateral variations of DAS, DMAS, EIBMV and EIBMV-DMAS using the images shown in Fig. 5 at the depths of (a) 7 $mm$ and (b) 11 $mm$.

## V. Discussion

The main enhancement obtained by the proposed algorithm is the lower level of sidelobes and higher resolution compared to DMAS and EIBMV. The low quality images, high effects of off-axis signals and high level of sidelobes, obtained with DAS, are mainly due to the blindness of DAS. On the other hand, DMAS beamformer is a non-linear algorithm and leads to high level of off-axis signals rejection due to its correlation process. In DMAS beamformer, a linear combination of the received signals is used to weigh the samples related to each elements of array. The resolution improvement by DMAS and the presence of noise in the reconstructed images degrade the quality of images. In EIBMV beamformer, an adaptive procedure is used, and samples are adaptively weighted to obtain a significant resolution improvement and low levels of sidelobe. However, the presence of noise is not degraded well enough in the reconstructed images using EIBMV method, seen in Fig. 3(c), Fig. 5(c). It should be noted that the frequency dependent attenuation of acoustic waves and optical attenuation of light should make the SNR lower in higher depths, as can be seen in TABLE I. However, the areas considered for $P_{signal}$ and $P_{noise}$ directly affect the SNRs. Thus, considering the way that we have used to calculate the SNRs, it is not possible to compare the SNRs at different depths. Since the same area are used for all the beamformers, we can have a fair comparison between the methods at each depth.

MV-based methods can be an appropriate choice when it comes to resolution. EIBMV is an algorithm which uses the eigenspace of the estimated covariance matrix to maintain the resolution of the MV algorithm and reduce the levels of sidelobes significantly compared to MV. However, the results show that the presence of noise still degrades the quality of the reconstructed PA images using EIBMV. There are multiple terms each of them can be interpreted as a DAS with different lengths of array in the expansion of DMAS algebra, leading to the low resolution of DMAS algorithm. It was proposed to used EIBMV instead of the terms in the mathematical expansion of DMAS in order to improve the resolution and level of sidelobes of the DMAS . However, as shown in (14), the number of contributing samples in each term of the expansion is different. The length of the subarray in the spatial smoothing highly effects the performance of the EIBMV algorithm, and in (14) there are some terms representing a low length of array and subarray. To address this problem, necessary terms are added to each term, and then, EIBMV algorithm is applied on it. Instead of a summation, interpreted as a DAS, EIBMV has been used in the introduced algorithm twice, to suppress the noise and sidelobes of EIBMV. Since the correlation procedure of DMAS contributes in the introduced method, the sidelobes and noise level of EIBMV are reduced. It should be noted that the proposed method in this paper would outperform DS-DMAS (presented in [18]). This is mainly due to the fact that EIBMV-DMAS uses the eigen-decomposition of covariance matrix, resulting in a higher noise suppression. In addition, weighting methods ( [22], [23]) can be applied on the proposed beamformer in order to further improve the PA image. In this paper, the EIBMV-DMAS has been evaluated numerically and experimentally. It should be noted that the processing time of the proposed method is higher than other mentioned beamformers. The correlation process of DMAS needs more time compared to DAS, and EIBMV needs more time to adaptively calculate the weights. EIBMV-DMAS uses two stages of EIBMV algorithm and a correlation procedure, so it is expected to have a higher processing time in comparison with EIBMV and DMAS. DAS, DMAS, EIBMV and the proposed method impose a complexity of $O(M)$, $O(M^2)$, $O(L^3)$ and $O(L^3)$, respectively. The level of sidelobes improvement and higher SNR obtained by the EIBMV-DMAS is visible in the reconstructed images. The proposed algorithm significantly outperforms DMAS and EIBMV in the terms of resolution and levels of sidelobes, respectively, mainly due to having the specifications of DMAS and EIBMV at the same time. The main drawback of the proposed method could be the higher computational burden in comparison with DMAS and EIBMV, but reducing the complexity of MV and EIBMV beamformers are extensively being investigated. Having MV-based methods with a lower computational burden would reduce the computational burden of the introduced method in this paper.

TABLE II: SNR ($dB$) Values at Different Depths for the Experimental PA Images.

| Depth($mm$) | DAS | DMAS | EIBMV | EIBMV-DMAS |
|---|---|---|---|---|
| 7 | 32.98 | 36.36 | 47.56 | 63.11 |
| 11 | 32.77 | 40.95 | 76.61 | 105.62 |

## VI. Conclusion

Expanding DMAS algebra results in several terms of DAS. In this paper, we proposed a novel beamforming algorithm based on the integration of EIBMV and DMAS algorithms, namely EIBMV-DMAS. The existing summation in the expansion of DMAS algebra (interpreted as DAS) was used, and it was proposed to employ the EIBMV beamforming instead of them. Introduced algorithm was evaluated numerically and experimentally. It was shown that EIBMV-DMAS beamformer improves the resolution, and reduces the level of noise and sidelobes compared to other concerned beamformers, but at the expense of higher computational burden. Quantitative comparison of the experimental images (at the depth of 11 $mm$) indicated that EIBMV-DMAS algorithm significantly reduces the SNR for about 162% and 38%, compared to DMAS and EIBMV, respectively.

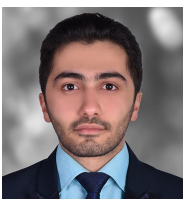

**Moein Mozaffarzadeh** received his BSc degree in electronics engineering from Babol Noshirvani University of Technology, Mazandaran, Iran, in 2015, and his MSc degree in biomedical engineering from Tarbiat Modares University, Tehran, Iran, in 2017. He is currently a research assistant at the Tarbiat Modares University, Tehran University of Medical Sciences, and the Research Center for Biomedical Technologies and Robotics, Institute for Advanced Medical Technologies, Tehran, Iran. His current research interests include photoacoustic image reconstruction, ultrasound beamforming, and biomedical imaging.

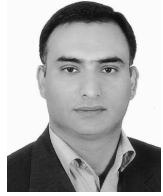

**Ali Mahloojifar** received his BSc degree in electronic engineering from Tehran University, Tehran, Iran, in 1988, and he received his MSc degree in digital electronics from Sharif University of Technology, Tehran, Iran, in 1991. He obtained his PhD in biomedical instrumentation from the University of Manchester, Manchester, United Kingdom, in 1995. He joined the Biomedical Engineering Group at Tarbiat Modares University, Tehran, Iran, in 1996. His research interests include biomedical imaging and instrumentation.

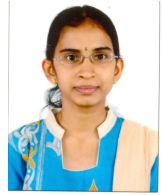

**Vijitha Periyasamy** is a project officer in the School of Chemical and Biomedical Engineering, Nanyang Technological University, Singapore. She has a bachelor's degree in medical electronics from Visvesvaraya Technological University, India. Her research interests include biomedical image processing for clinical evaluation, development of multimodal imaging systems, Monte Carlo simulation for light transport in biological tissue, and application of bioengineering for different medical practice systems.

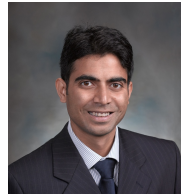

**Manojit Pramanik** received his PhD in biomedical engineering from Washington University in St. Louis, Missouri, USA. He is currently an assistant professor of the School of Chemical and Biomedical Engineering, Nanyang Technological University, Singapore. His research interests include the development of photoacoustic/thermoacoustic imaging systems, image reconstruction methods, clinical application areas such as breast cancer imaging, molecular imaging, contrast agent development, and Monte Carlo simulation of light propagation in biological tissue.

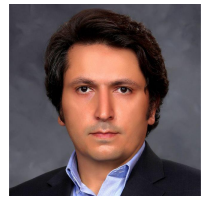

**Mahdi Orooji** received the BSc degree in Electrical Engineering from the University of Tehran, Iran, in 2003, and the MSc. and PhD. in Communication Systems and Signal Processing from the School of Electrical Engineering and Computer Science, Louisiana State University, Baton Rouge, Louisiana, USA in 2010 and 2013, respectively. From 2013 to 2016, he was a Postdoctoral Fellow in the Center for Computational Imaging and Personalized Diagnostics, Case Western Reserve University, Cleveland, Ohio, USA. Since 2016, he has been with Tarbiat Modares University, Tehran, Iran, where he is currently an Assistant Professor of Bioelectrical Engineering. His main areas of research interest include the computer aided diagnostics system, artificial intelligence and machine vision in medical images.